\documentstyle[12pt]{article}

\begin{document}

\begin{titlepage}

\begin{center}
\baselineskip 36pt
{\LARGE {\bf 't Hooft's Order-Disorder Parameters and the Dual Potential}}\\
\vspace{1cm}
\baselineskip 16pt
{\large CHAN Hong-Mo}\footnote{chanhm\,@\,v2.rl.ac.uk}\\
\vspace{.2cm}
{\it Rutherford Appleton Laboratory,\\
    Chilton, Didcot, Oxon, OX11 0QX, United Kingdom}\\
\vspace{.5cm}
{\large TSOU Sheung Tsun}\footnote{tsou\,@\,maths.ox.ac.uk}\\
\vspace{.2cm}
{\it Mathematical Institute, Oxford University,\\
    24-29 St. Giles', Oxford, OX1 3LB, United Kingdom}\\
\end{center}

\vspace{1cm}
\begin{abstract}
It is shown that the operator $B(C) = {\rm Tr} [P \exp i \tilde{g} \oint 
\tilde{A}_i(x) dx^i]$ constructed with the recently derived dual 
potential $\tilde{A}(x)$ and a coupling $\tilde{g}$ related
to $g$ by the Dirac quantization condition satisfies the correct 
commutation relation with the Wilson operator ${\rm Tr} [P \exp ig \oint 
A_i(x) dx^i]$ as required by 't Hooft for his order-disorder parameters.
\end{abstract}

\end{titlepage}

\clearpage

\baselineskip 16pt

In his study of the confinement problem in nonabelian gauge theories,
't Hooft introduced 2 loop-dependent operators $A(C)$ and $B(C)$ with
the following commutation relations: \cite{thooft}
\begin{equation}
A(C) B(C') = B(C') A(C) \exp (2 \pi i n/N)
\label{thooftcom}
\end{equation}
for $su(N)$ gauge symmetry, and any two spatial loops $C$ and $C'$ with
linking number $n$ between them.  $A(C)$ was explicitly given as:
\begin{equation}
A(C) = {\rm Tr} \left[ P \exp ig \oint_C A_i(x) dx^i \right],
\label{Wilsonloop}
\end{equation}
and in the words of 't Hooft measures the magnetic flux through $C$
and creates electric flux along $C$.  On the other hand, $B(C)$ 
measures the electric flux through $C$ and creates magnetic flux along
$C$, and plays thus an exactly dual role to $A(C)$.  For lack of a
dual potential, however, $B(C)$ was not given a similar explicit 
expression.  

In a recent paper \cite{Chanftsou}, it was shown that a dual potential 
$\tilde{A}_\mu(x)$ does exist in nonabelian gauge theories, and the 
explicit though complicated transform between dual variables is given. 
That being so, one ought to have:
\begin{equation}
B(C) = {\rm Tr} \left[ P \exp i \tilde{g} \oint_C \tilde{A}_i(x) dx^i \right]
\label{Wilsonloopt}
\end{equation}
as the explicit expression for $B(C)$, with the dual coupling $\tilde{g}$
related to $g$ by a Dirac quantization condition.  The purpose of the 
present note is to show that this is indeed the case.

Recall first the proof of (\ref{thooftcom}) for abelian fields.  In that
case, one can ignore the trace and the ordering in (\ref{Wilsonloop}) and
(\ref{Wilsonloopt}) so that $A(C)$ and $B(C)$ are genuine exponentials
of line integrals.  Using Stokes' theorem, one of the exponents, say of
$B(C')$, can be written as a surface integral, thus:
\begin{equation}
B(C') = \exp - i \tilde{e} \int\int_{\Sigma_{C'}} \mbox{}^*\!F_{ij} 
   d \sigma^{ij},
\label{BCprime}
\end{equation}
or, by the definition of the Hodge star (dual transform)
\begin{equation}
\mbox{}^*\!F_{\mu\nu} = - \frac{1}{2} \epsilon_{\mu\nu\rho\sigma} 
   F^{\rho\sigma},
\label{Hodgestar}
\end{equation}
in terms of the electric field strength ${\cal E}_i = F_{0i}$ as:
\begin{equation}
B(C') = \exp i \tilde{e} \int\int_{\Sigma_{C'}} {\cal E}_i d \sigma^i,
\label{BCpp}
\end{equation}
where $\Sigma_{C'}$ is some surface both spanning over and bounded by $C'$.  

Consider first the simple case for linking number 1 between $C$ and $C'$.
The loop $C$ in that case will intersect the surface $\Sigma_{C'}$ at some 
point $x_0$.  ($C$ may of course intersect $\Sigma_{C'}$ more than once, 
but the extra intersections occurring pairwise with opposite orientations, 
their contributions to the commutator will all cancel, leaving in effect
just one intersection.)  Except at this point $x_0$, all points on $C$ 
are spatially separated from points on $\Sigma_{C'}$ so that, using the
canonical commutation relation between $A_i(x)$ and ${\cal E}_i(x)$:
\begin{equation}
[{\cal E}_i(x), A_j(x')] = i \delta_{ij} \delta(x-x')
\label{eacomm}
\end{equation}
we have:
\begin{equation}
\left[ ie \oint_C A_i(x) dx^i, i \tilde{e} \int \!\! \int_{\Sigma_{C'}}
   {\cal E}_j d \sigma^j \right] = i e \tilde{e},
\label{expcomm}
\end{equation}
a c-number.  Hence we conclude that 
\begin{equation}
A(C) B(C') = B(C') A(C) \exp i e \tilde{e}
\label{abeliancom}
\end{equation}
which by the Dirac quantization condition:
\begin{equation}
e \tilde{e} = 2 \pi
\label{diraccond}
\end{equation}
gives the answer (\ref{thooftcom}) for $n = 1$ as required.  In case
$C'$ winds around $C$ more than once, say $n$ times, then $C$ will
intersect $\Sigma_{C'}$ at effectively $n$ points for each of which 
the above applies, so that (\ref{thooftcom}) still remains valid.

What happens when we generalize to the nonabelian case? Then $A(C)$
and $B(C)$ are each a trace of an ordered product of noncommuting
factors for which no Stokes' Theorem applies.  Nevertheless, one finds
that one may still associate with each a surface in an analogous fashion.
Take $B(C')$, for example.  The phase factor:
\begin{equation}
\tilde{\Phi}(C') = P \exp i \tilde{g} \oint_{C'} \tilde{A}_i dx^i
\label{PhiCp}
\end{equation}
of which $B(C')$ is the trace, can be written, according to 
ref. \cite{Chanftsou}, as:
\begin{equation}
\tilde{\Phi}(C') = \prod_{t = 0 \rightarrow 2 \pi} (1 - i \tilde{g} 
   \tilde{W}[\eta|t]) \sim \prod_{t = 0 \rightarrow 2 \pi}
   \exp - i \tilde{g} \tilde{W}[\eta|t],
\label{PhiCpp}
\end{equation}
where $\eta$, for $t = 0 \rightarrow 2 \pi$, is a parametrization of
$C'$, and:
\begin{equation}
\tilde{W}[\eta|t] = \int_{\eta_0}^{\eta(t)} \delta \eta^{' \nu}(t)
   \tilde{E}_\nu[\eta'|t],
\label{Wtildeint}
\end{equation}
$\tilde{E}_\nu[\eta|t]$ and $\tilde{W}[\eta|t]$ being both `segmental'
quantities depending on a segment of $C'$ around the point $\eta(t)$
on it.  These `segmental' quantities were used to establish dual symmetry
for nonabelian theories in ref. \cite{Chanftsou}, to which the reader is
referred for detailed explanation of their significance.  We note here 
only that the integral in (\ref{Wtildeint}) denotes a `segmental' integral
along some path from a reference point $\eta_0$ to the point $\eta(t)$,
so that in ordinary space, this path appears as a ribbon.  Piecing 
such ribbons together as $\eta(t)$ moves along $C'$ in (\ref{PhiCpp}),
one obtains a surface $\Sigma_{C'}$ spanning over and bounded by $C'$
as suggested.  In as much as the reference point $\eta_0$ and the path
joining it to $\eta(t)$ are both arbitrary for (\ref{Wtildeint}) to
hold, one can choose $\Sigma_{C'}$ to be completely space-like.  
This surface will again intersect the loop $C$ of $A(C)$ at some point
$x_0$.  (Previous remarks in the abelian case about multiple intersections 
and higher linking numbers between $C$ and $C'$ will still apply and will 
not be repeated.) 

To proceed further, write for convenience:
\begin{equation}
A(C) =  {\rm Tr} \left[ \prod_{s = 0 \rightarrow 2 \pi} \phi(s) \right],
\label{ACprod}
\end{equation}
\begin{equation}
B(C') = {\rm Tr} \left[ \prod_{t = 0 \rightarrow 2 \pi} \tilde{\phi}(t) \right],
\label{BCprod}
\end{equation}
with
\begin{equation}
\phi(s) = \exp i g A_i(\xi(s)) \dot{\xi}^i(s) ds,
\label{phis}
\end{equation}
\begin{equation}
\tilde{\phi}(t) = \exp i \tilde{g} \int_{\eta_0}^{\eta(t)} \delta
   \eta^{'i}(t) \tilde{E}_i[\eta'|t],
\label{phit}
\end{equation}
where all products are meant to be properly ordered.  Hence,
\begin{equation}
A(C) = {\rm Tr} [V \phi(s_0)],
\label{ACV}
\end{equation}
\begin{equation}
B(C') = {\rm Tr} [\tilde{\phi}(t_0) \tilde{V}],
\label{BCV}
\end{equation}
for
\begin{equation}
V = \prod_{s = 0 \rightarrow s_0} \phi(s) \prod_{s = s_0 \rightarrow 2 \pi}
   \phi(s),
\label{V}
\end{equation}
\begin{equation}
\tilde{V} = \prod_{t = 0 \rightarrow t_0} \tilde{\phi}(t)
   \prod_{t = t_0 \rightarrow 2 \pi} \tilde{\phi}(t),
\label{Vtilde}
\end{equation}
where $s_0$ and $t_0$ refer to the intersection of $C$ with $\Sigma_{C'}$,
namely such that $\xi(s) = x_0$, and that the ribbon for $t = t_0$,
as defined by (\ref{phit}), passes through $x_0$.  To avoid being entangled
with the somewhat extraneous noncommutativity of these quantities 
which is due to their being elements of the gauge algebra, we
rewrite them in terms of their internal symmetry components, thus:
\begin{equation}
A(C) = \sum_{a,b} V_{ab} \phi_{ba}, \ a,b = 1,.... N,
\label{ACincomp}
\end{equation}
\begin{equation}
B(C') = \sum_{c,d} \tilde{\phi}_{cd} \tilde{V}_{dc}, \ c,d = 1, ..., N,
\label{BCincomp}
\end{equation}
where the indexed quantities are now just c-numbers in internal symmetry
space, though still operators in the quantum mechanical Hilbert space.
We note, however, that except for the pair $\phi_{ab} = \phi_{ab}(s_0)$
and $\tilde{\phi}_{cd} = \tilde{\phi}_{cd}(t_0)$, all other factors are
spatially separated and would thus mutually commute.  That being the case, 
then supposing we assume that:
\begin{equation}
\phi_{ba} \tilde{\phi}_{cd} = \tilde{\phi}_{cd} \phi_{ba} \exp 2 \pi i/N,
\label{phicom}
\end{equation}
we would obtain (\ref{thooftcom}) for $n=1$ as desired.

Let us examine then the relation (\ref{phicom}).  It would be valid if 
the exponents of $\phi$ and $\tilde{\phi}$ in (\ref{phis}) and (\ref{phit})
satisfy the following commutation relation:
\begin{equation}
\left[ i g A_i(\xi(s_0)) \dot{\xi}^i(s_0) ds_0, i \tilde{g} \int_{\eta_0}
   ^{\eta(t_0)} \delta \eta^{'j}(t_0) \tilde{E}_j[\eta'|t_0] \right] 
   = 2 \pi i/N.
\label{expcom}
\end{equation}
We note that in (\ref{expcom}), although $A_i$ and $\tilde{E}_j$ are
both matrices in internal symmetry indices, they are to be regarded as
elements of different algebras, since in (\ref{phicom}) these indices
are not summed.  In other words, if we write:
\begin{equation}
A_i(x) = A_i^\alpha X_\alpha,
\label{Aicomp}
\end{equation}
\begin{equation}
\tilde{E}_i[\eta|t] = \tilde{E}_i^\alpha[\eta|t] \tilde{X}_\alpha,
\label{Eitcomp}
\end{equation}
where although $X_\alpha$ and $\tilde{X}_\alpha$ are matrices representing
the generators of the same gauge Lie algebra, they are to be regarded as
corresponding to different degrees of freedom, like isospins of different
particles.  They therefore commute as far as (\ref{expcom}) is concerned,
their product there being the tensor product, not the ordinary matrix
product.

To obtain (\ref{phicom}), we write, recalling the dual transform defined
in ref. \cite{Chanftsou}:
\begin{equation}
\tilde{E}_i[\eta|t] = - \frac{2}{\bar{N}} \epsilon_{ij\rho\sigma}
   \dot{\eta}^j(t) \int \delta \xi ds \, \omega(\xi(s)) E^\rho[\xi|s]
   \omega^{-1}(\xi(s)) \dot{\xi}^\sigma(s) \dot{\xi}^{-2}(s) 
   \delta(\xi(s)-\eta(t)),
\label{dualtransf}
\end{equation}
where using arguments similar to those given there for showing the abelian
reduction of (\ref{dualtransf}) to the Hodge star we can rewrite the 
right-hand side as:
\begin{equation}
-\frac{1}{\bar{N}} \epsilon_{ijk0} \dot{\eta}^j(t) \! \int \! \delta \xi ds \,
   \omega(\xi(s)) U_\xi(s) F^{k0}_\alpha(\xi(s)) \tilde{X}^\alpha
   U^{-1}_\xi(s) \omega^{-1}(\xi(s)) \dot{\xi}^{-2}(s) \delta(\xi(s) - \eta(t)),
\label{rhside}
\end{equation}
with $U_\xi(s)$ being an element of the gauge group.  Then, using the 
canonical commutation relation for ${\cal E}^\alpha_i = F^\alpha_{0i}$:
\begin{equation}
[{\cal E}^\alpha_i(x), A^\beta_j(x')] = i \delta^{\alpha\beta} \delta_{ij} 
   \delta(x-x'),
\label{Canoncom}
\end{equation}
valid in the temporal gauge $A_0^\alpha = 0$, we obtain for the commutator
in (\ref{expcom}):
\begin{equation}
i \frac{g \tilde{g}}{\bar{N}} \omega(x_0) \left[ \int \delta \xi 
   U_\xi(s_0) X^\alpha \tilde{X}_\alpha U^{-1}_\xi(s_0) \dot{\xi}^{-2}(s_0)
   \delta(\xi(s_0) - x_0) \right] \omega^{-1}(x_0).
\label{expcomrh}
\end{equation}
The quantity $X^\alpha \tilde{X}_\alpha$, however, is just a number
\cite{Ross}:
\begin{equation}
\sum_\alpha X^\alpha_{ab} \tilde{X}^\alpha_{cd} = \frac{1}{2} \left[
   \delta_{ad} \delta_{cb} - \frac{1}{N} \delta_{ab} \delta_{cd} \right].
\label{sunross}
\end{equation}
Hence, the integral in (\ref{expcomrh}) can be done and cancels the 
normalisation factor $\bar{N}$ in the denominator as defined in
ref. \cite{Chanftsou} giving for the commutator in (\ref{expcom}) just:
\begin{equation}
i g \tilde{g} \left(\pm \frac{1}{2} - \frac{1}{2N} \right)
\label{excomrhp}
\end{equation}
for respectively the states symmetric or antisymmetric under the 
interchange of tilde with no tilde.  We would then obtain the
desired result in (\ref{expcom}) if $g$ and $\tilde{g}$ satisfy the
Dirac quantization condition:
\begin{equation}
g \tilde{g} = 4 \pi.
\label{Diraccond}
\end{equation}

That this condition holds in the standard normalization convention 
adopted here\footnote{Notice that this is a different convention
from that adopted in our earlier publications on the subject, e.g.
\cite{Chanftsou}; hence the different form of the Dirac condition.},
where the action is written as:
\begin{equation}
{\cal A} = -\frac{1}{4}\int d^4x{\rm Tr}[\tilde{F}_{\mu\nu} \tilde{F}^{\mu\nu}]
   + \int d^4x \bar{\tilde{\psi}} (i \partial_\mu \gamma^\mu - m) \tilde{\psi}
\label{Action}
\end{equation}
can be seen by writing
\begin{equation}
\tilde{D}^\mu \tilde{F}_{\mu\nu} = - \tilde{g} (\bar{\tilde{\psi}}
   \gamma_\nu \tilde{X}^\alpha \tilde{\psi}) \tilde{X}_\alpha
\label{dualcurrent}
\end{equation}
as a dual current in the direct (i.e. no tilde) description in terms
of the loop space curvature $G_{\mu\nu}[\xi|s]$.  For consistency with
obtaining the correctly quantized monopole charge, we need:
\begin{equation}
\exp \left[ -ig \tilde{g} \epsilon_{\mu\nu\rho\sigma} (\bar{\tilde{\psi}}
   \gamma^\rho \tilde{X}^\alpha \tilde{\psi}) \dot{\xi}^\sigma
   \tilde{X}_\alpha \delta \xi^\mu \delta \xi^\nu \right] = \exp 2 \pi i/N.
\label{quantcond}
\end{equation}
This condition being invariant, that it is satisfied by (\ref{Diraccond})
can easily be checked by giving $\tilde{\psi}$ a specially simple
orientation, say $(1,0,...,0)$, and using the standard representations
for the $su(N)$ matrices $\tilde{X}_\alpha$.

With the Dirac condition (\ref{Diraccond}) now established, the `proof' of
the validity of 't Hooft's commutation relation (\ref{thooftcom}) for
the operators $A(C)$ and $B(C)$ in (\ref{Wilsonloop}) and (\ref{Wilsonloopt})
is then complete, although for lack of a general calculus for handling loop 
operations which is keenly felt throughout the scheme of ref. \cite{Chanftsou}, 
the `proof' is of necessity not as rigorous as one could desire.

\end{document}